\documentstyle[12pt]{article}

\newcommand{\ket}[1]{\mbox{$\vert #1 \rangle$}}

\newcommand{\beq}{\begin{equation}}
\newcommand{\eeq}{\end{equation}}
\newcommand{\bea}{\begin{eqnarray}}
\newcommand{\eea}{\end{eqnarray}}
\newcommand{\ba}{\begin{array}}
\newcommand{\ea}{\end{array}}

\begin{document}

\begin{center}
{\large
 The Kondo-Hubbard model at half-filling\hfill}
\end{center}

\vspace{1cm}
\mbox{}\\
 J.~P\'erez-Conde$^a$, F.~Bouis$^b$ and P.~Pfeuty$^b$. \\

\mbox{} $a$- Depto. \ de F{\'\i}sica, Universidad
 P\'ublica de Navarra, E-31006, Pamplona, Spain  \\
\mbox{} $b$- Laboratoire L\'eon Brillouin - CE-Saclay 91191,
Gif-sur-Yvette, France \\

\vspace{1cm}
 We have analyzed the antiferromagnetic ($J>0$) Kondo-Hubbard lattice with the band
 at half-filling by means of a perturbative approach in the strong coupling limit, the small parameter is
 an arbitrary tight-binding band. The results are valid
 for any band shape and any dimension. We have obtained the energies of
 elementary charge and spin excitations as well as the magnetic correlations
 in order to elucidate the magnetic and
 charge behaviour of the Kondo lattice at half-filling. Finally, we have
 briefly analyzed the ferromagnetic case 
 ($J<0$), which is shown to be equivalent to an effective antiferromagnetic Heisenberg model.

\newpage

 We consider here the Kondo lattice model first introduced to
 describe heavy-fermion systems \cite{do77}, but studied here for the 
 particular
 half-filled case with one band electron per site so that the model can 
 describe an insulator with Kondo like properties \cite{af92}.
 The Hamiltonian consists in a periodic lattice of magnetic atoms 
in a metallic background modelled by $f$ orbitals and a tight-binding band respectively.
 The two systems interact through an exchange coupling constant $J$ between
 the spin density of
 the band electrons and the spin of the magnetic atoms. 
Additionally, we introduce a Coulomb intra-site repulsion between the
 electrons of strenght $U$.                
The model Hamiltonian is,
\[
H=\sum_{i,j,s}t_{ij}c_{is}^{\dagger}c_{js}+
J\sum_{i}{\bf s}_{c i}\cdot{\bf S}_{f i}+
U\sum_{i}c_{i\downarrow}^{\dagger}c_{i\downarrow}
c_{i\uparrow}^{\dagger}c_{i\uparrow}
=H_{t}+H_{0}.
\]
The first term is the tight-binding band where $t_{ij}=t_{ji}$ is an 
unrestricted hopping parameter ($t_{ii}=0$). The second term is the spin-spin
 interaction where 
 ${\bf s}_{c i}=\frac{1}{2}c_{is}^{\dagger}{\bf \sigma}_{s s^{'}}c_{is}$
is the spin density of the conduction electrons, ${\sigma^{a}}$ are the Pauli 
matrices and ${\bf S}_{fi}$ ($S_{f}=1/2$) is the spin of the magnetic
 impurities. 

 In the $t_{ij}=0$ limit the sites are 
 uncorrelated and we have just to obtain the eigenvalues and eigenvectors of 
 the one-site Hamiltonian 
 $h_{0i}=J{\bf s}_{c i}\cdot{\bf S}_{f i}+U n_{i\downarrow}
 n_{i\uparrow}$. We obtain the complete solution, eigth eigenstates:  
 a singlet 
 state \ket{s_{i}}
 with energy $-3J/4$, two doublets $\{\ket{d_{0i}^{\pm}}, \ket{d_{2i}^{\pm}}\}$
 with energy $0$ and $U$ respectively, and finally a triplet $\{\ket{t_{i}^{0}},
 \ket{t_{i}^{\pm}}\}$ with energy $J/4$. 

 From the one site solution we can build and classify by inspection 
 all the eigenstates of the complete Hamiltonian $H_{0}$. We restrict the 
 analysis to the half-filled band in any dimension. The ground state in this 
 limit is a non degenerate singlet which describes an insulator without any
 magnetic structure. This limit seems appropiate to describe qualitatively some
 of the properties of the FeSi \cite{af92} at zero temperature.

 When we introduce the band term the sites are correlated and the physical 
 picture can be modified. We are interested in the magnetic and insulating
 behavior of the system and therefore we have obtained
 the spin and charge energy gaps and the spin-spin correlations.
 The spin gap is defined as the difference between the energies
 of the ground state and the triplet state, 
$\Delta_{S}= E_{0}(N,S=1)-E_{0}(N,S=0)$. 
 The zero order ($t_{ij}=0$) value of $\Delta_{S}$ is $J$ and it corresponds
 to the difference between the energy of the first excited state, a triplet,
  and the singlet state.
 The triplet is $N$-fold degenerate in absence of hopping. Hence, we need to
 apply the perturbation formalism corresponding to the degenerate case, then
 obtain an effective Hamiltonian in the $S=1$ subspace (see \cite{pp95} and
 \cite{jamais} for details) and finally, we get the following expression for
 the spin excitation energy,
\begin{equation}
\Delta_{S}{(\bf q)}=J+ (-2B_{1}^{S=1}+2B_{2}^{S=1})\sigma_{1}
-2B_{2}^{S=1}{\tau}^{(2)}({\bf q}).
\label{gap_s}
\end{equation}
where, 
$
\sigma_{1}=\int
\frac{d^{d}{\bf q}}{(2\pi)^{d}}
[\tau{({\bf q})}]^{2},
\quad
\tau^{(2)}{({\bf q})}=
\int
\frac{d^{d}{\bf q}^{\prime}}{(2\pi)^{d}}
\tau{({\bf q}^{\prime}+{\bf q})}
 \tau{({\bf q}^{\prime})}
$, 
 $B_{1}^{S=1}=-1/(3J+U)$ and  $B_{2}^{S=1}=-1/(J+2U)$ and $\tau{({\bf q})}$
 is the band dispersion.

 This expression is ${\bf q}$-dependent. We could then define the absolute gap 
 as the minimal value of (\ref{gap_s}), $\Delta_{S}({\bf q}_{min})$, which 
 corresponds to a value of the wave vector ${\bf q}_{min}$ which minimizes 
 the energy. The nearest-neighbor case gives,
$\Delta_{S}({\bf q})=J-20t^{2}(3J)d$.
 From this result on one hand one could think that there exist 
 a phase transition between the triplet and the singlet state. On the other hand
 it is believed that the half-filled Kondo lattice in one dimension at zero 
 temperature is a spin liquid  \cite{th92,sn96}, so that 
 the spin gap is non zero for any $J\neq 0$. The contradiction is due, of 
 course, to the poor approximation to the true gap we have obtained. 
 Nevertheless, if we accept the existence of the spin gap, we can deduce that 
 our result is qualitatively good even for moderate values of $J/t$ (at $J=3$ 
 the gap is always positive, so there is no transition). 
 When more terms in the series in $t/J$ for the spin gap will be available, 
 an analysis of the series should tell us about the critical value $(t/J)_{c}$
 at which the spin gap closes and a magnetic phase transition takes place. 
 Another interesting point is that even when $J$ is small and $U$ is sufficiently
 large we get a non zero spin gap so that the Kondo insulator behavior could be 
 obtained in a Kondo lattice system with small Kondo coupling if the conduction
 electrons are by themself strongly correlated: when $J\sim 0$ then
 $B_{1}^{S=1}=B_{2}^{S=1}=-1/(2U)$, so that $\Delta_{S}\sim t^2/U >0$.
 The spin-spin correlations between the magnetic atoms have been also obtained
 for the nearest-neighbor hopping, 
$\langle S_{fi}\cdot S_{fj}\rangle =-(5/6)(t/ J)^2$
(we write the result in the simple case U=0).

 The insulating gap is defined as,
$\Delta_{I}=\mu^{+}-\mu^{-}=E_{0}(N+1,S=1/2)+E_{0}(N-1,S=1/2)-2E_{0}(N,S=0)
$. 
  It can be seen as the difference between the chemical potential with $N+1$
 particles, $\mu^{+}$ and $\mu^{-}$, the chemical potential at $N-1$ particles. 
 With this definition, the gap is the second 
 derivative of the energy with respect to the number of particles. In this sense the 
 insulator gap could be considered as an order parameter of the insulator-metal
 transition. 
 The total energy dispersion in the $N\pm 1$ subspace is,
\beq
E_{0}^{N \pm 1}{(\bf q)}=-(N-1)\frac{3}{4}J+U
\pm\frac{1}{2}\tau ({\bf q})+
3(B_{2}^{N\pm 1}-B_{1}^{N\pm 1})\sigma_{1}+2NB_{1}^{N\pm 1}\sigma_{1}-
 B_{1}^{N\pm 1}\tau^{2}{(\bf q)},
\label{en_np1}
\eeq
 where 
$B_{1}^{N\pm 1}=B_{1}^{S=1}/2$, 
$B_{2}^{N\pm 1}=-1/(4J)$. 
 The gap can be now explicitly written. It is interesting, 
 however, to discuss the meaning of two different expressions for the gap.
 The first one is the ${\bf q}$-dependent generalization,
\beq 
\Delta_{I}=\frac{3 J}{2} +
 6(B_{2}^{N\pm 1}-B_{1}^{N\pm 1})\sigma_{1}-
 2B_{1}^{N\pm 1} \tau^{2}({\bf q}),
 \label{indirectgap}            
\eeq 
 where no first order term appear. The second expression of the gap can be 
 obtained if we strictly follow the prescription given in  (\ref{en_np1}),
 where the lowest energy value on the $N+1$ subspace and the
 highest energy value on the $N-1$ subspace are taken,
 \beq
\Delta_{I}=\frac{3 J}{2}+U
 -\frac{1}{2} (
 \tau ({\bf q}_{min}^{+})
 -\tau ({\bf q}_{max}^{-})
 ) +
 6(B_{2}^{N\pm 1}-B_{1}^{\pm})\sigma_{1}-
 B_{1}^{N\pm 1}(
 \tau^{2}({\bf q}_{min}^{+})+
 \tau^{2}({\bf q}_{max}^{-})
 )
 \label{directgap}
\eeq
 Both expressions are physically acceptable if one thinks, for example, in terms
 of excitations which can be created by direct or indirect absortion of a 
 photon. The former will be piloted by (\ref{directgap}) where the particle 
 momentum does not change and the later by (\ref{indirectgap}). 
 The chemical potential expressions can be then taken as effective 
 bands of the original Hamiltonian. If we explicitly write them as
 functions of $q$ in one dimension and  nearest-neighbor hopping
 (we take henceforth $U=0$ in order to simplify) we obtain 
 $ \mu^{+}=3J/4-t\cos{q}
          -t^{2}/(2J)
          +2t^{2}\cos^{2}{q}/(3J)$, 
 $\mu^{-}=-3J/4-t\cos{q}
          +t^{2}/(2J)
          -2 t^{2}\cos^{2}{q}/(3J)$.

 The direct gap (\ref{directgap}) at each value of $J/t$, is graphically given 
 by the difference between the two chemical potentials as functions of $q$.
 The indirect gap is the energy difference between the absolute minimum of the 
 $\mu^{+}$ band and the absolute maximum of the $\mu^{-}$ band.

 We have also considered the ferromagnetic interaction $J<0$. In this 
 case when the hopping term is zero the ground state is degenerate and there
 is a triplet $\{ t^{0}$, $t^{+}$, $t^{-}\}$ on each site. When the band term is
 included the low energy excitations are described by an (insulating)
 antiferromagnetic
  Heisenberg Hamiltonian, $H=-({2t^{2}}/{J}){\bf S}_{i}\cdot{\bf S}_{j}$, where
 ${\bf S}_{i}$ represents the spin one on site $i$.

This work was supported in part by spanish DGICyT under 
 contract no. PB95-0797.

\end{document}